\documentclass[structabstract]{aa}  
\usepackage{graphicx}
\usepackage{txfonts}
\usepackage[utf8]{inputenc}
\usepackage{times}
\usepackage{natbib}

\bibpunct{(}{)}{;}{a}{}{,}

\begin{document}
 \title{First evidence for a gravitational lensing-induced echo in gamma rays with Fermi LAT}

   \author{A.~Barnacka \inst{2,1}
          \and
          J-F.Glicenstein \inst{1}  
          \and 
          Y.~Moudden \inst{1}
          }

   \institute{DSM/IRFU, CEA/Saclay, F-91191 Gif-sur-Yvette, France
         \and
             Nicolaus Copernicus Astronomical Center, Warszawa, Poland\\
             %\email{abarnack@camk.edu.pl}
             }
  \offprints{glicens@cea.fr,abarnack@camk.edu.pl}
  \abstract
   {}
  { This article shows the first evidence for gravitational lensing phenomena 
    in high energy gamma-rays. This evidence comes from the observation of a
gravitational lens induced echo in the light curve of the distant blazar PKS 1830-211.}
   {
Traditional methods for the estimation of time delays in gravitational lensing
systems rely on the cross-correlation of the light curves of the individual 
images. In this paper, we use 300 MeV-30 GeV photons 
detected by the Fermi-LAT instrument.
The Fermi-LAT instrument cannot separate the images of known lenses.
The observed light curve is thus the superposition of individual 
image light curves.
The Fermi-LAT instrument has the advantage of providing long, evenly spaced, 
time series. In addition, the photon noise level is very low.      
This allows to use directly Fourier transform methods. 
   }
   { 
    A time delay between the two compact images of PKS 1830-211 has been 
searched for  both by the autocorrelation method and the 
``double power spectrum'' method. The double power spectrum shows a
3 $\sigma$ evidence for a time delay  of 27.5$\pm$1.3 days, consistent 
with the result from Lovell et al. (1998). The relative uncertainty on the 
time delay estimation is reduced from 20\% to 5\%. 
   }
   {}

   \maketitle
%
%________________________________________________________________

\section{Introduction}

The precise estimation of the time delay between components of lensed Active Galactic Nuclei (AGN) is crucial 
for modeling the lensing objects. In turn, more accurate lens models give better constraints on the Hubble constant.
More than 200 strong lens systems have been found. 
Most of them were discovered in recent years  by dedicated surveys such as 
 the Cosmic Lens All-Sky Survey \citep{2003Myers,2003Browne}
and the Sloan Lens ACS Survey \citep{2004Bolton}. 
The launch of the Fermi sattelite \citep{2009Atwood} in 2008 gives 
the opportunity
to observe gravitational lensing phenomena with high energy gamma rays. 
The observation strategy of Fermi-LAT, which surveys 
the whole sky in 190 minutes, allows a regular sampling of quasar light curves with 
a period of a few hours.
Since the launch of the Fermi satellite, the LAT  instrument has been 
collecting high energy photons for  more than 800 days. 
This paper deals with the observation and estimation of time delays of strong 
lensing systems and not with the detection and use of microlensing phenomena as in \citet{2003Torres}.

The multiple images of a gravitational lensed AGN cannot be  
directly observed with the high energy gamma-ray instruments such as  
the Fermi-LAT, Swift or ground based Cerenkov telescopes, 
due to their limited angular resolutions. The angular resolution of 
these instruments is at best of a few arcminutes (in the case of HESS), 
when the typical separation of the images for quasar lensed by galaxies is a few arcseconds.
Since the multiple images cannot be observed, methods based on the detection
of an echo in the light curve are studied in this paper.
The observed signal is the superposition of the intrinsic light curve of the AGN
and at least one delayed copy of this light curve, with a different amplitude.

The method has been tested on simulated light curves and on Fermi observations 
of the very bright radio quasar PKS 1830-211, for which the 
time delay 
has been accurately measured 
by \citet{1998Lovell} using radio observations.   
The paper is organized as follows: we first give a very brief summary of properties of PKS 1830-211 and of Fermi LAT data towards this AGN.
Then we introduce the methods for time delay estimation. The last section is devoted to the measurement of the time delay 
between the two compact components of PKS 1830-211. 

%________________________________________________________________
\section{The PKS 1830-211 gravitational lens }
The AGN PKS 1830-211 is a variable, bright radio source and an X-ray blazar. 
Its redshift was measured to be z=2.507 \citep{1999Lidman}. 
The blazar was detected in the $\gamma$-ray wavelengths with EGRET. 
The association of the EGRET source
with the radio source was done by \cite{1997Mattox}. 
The classification of PKS 1830-211 as a gravitational lensed quasi-stellar object
 was first proposed by \cite{1988Rao}.

Two potential lenses have been identified along the line of sight to PKS 1830-211
by the observation of molecular absorption at centimeter and millimeter wavelengths.
One possible lensing object is a galaxy 
at redshift  z=0.19 \citep{1996Lovell}, and the other a galaxy 
at redshift z=0.89 \citep{1996Wiklind}.
According to \citet{2002Winn}, the actual lens is the galaxy at redshift z=0.89.

PKS 1830-211 is observed in radio as an elliptical ring-like structure connecting 2 bright sources distant of roughly one
arcsecond \citep{1991Jauncey}.
The compact components were separately observed 
by the Australia Telescope Compact Array at 8.6 GHz for 18 months. 
These observations and the subsequent analysis made by \cite{1998Lovell} 
gave a magnification ratio between the 2 images equal to $1.52\pm 0.05$ 
and a time delay of $26^{+4}_{-5}$ days. 
%----------------------------------------------------------- lc - Light Curve -----------
   \begin{figure*}
   \centering
 \includegraphics[width=15cm,angle=0]{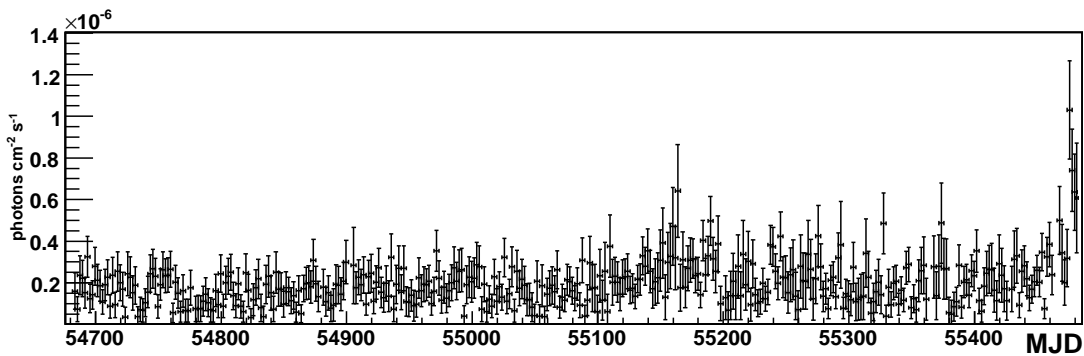} \\
 \caption{Fermi LAT light curve of PKS 1830-211,  with a 2 days binning. The energy range is 300 MeV to 300 GeV.} 
         \label{lc}
   \end{figure*}
%-----------------------------------------------------------------------------

\subsection{Fermi LAT data on PKS 1830-211}
PKS 1830-211 has been detected by the Fermi-LAT instrument with a detection significance above 41 Fermi Test Statistic (TS), 
equivalent to a 6 $\sigma$ effect \citep{2010Abdo}. The long-term light curve is presented on figure \ref{lc} with a 2 days binning.
The data analysis from this paper uses a 2 days binning, which provides 
a sufficient photon statistic per bin with a time span per bin
much shorter than 28 days. The data analysis was cross-checked 
by binning the light-curve into 1 day and 23 hours bins, with
similar results.  The data were taken between August 4 2008 and 
October 13 2010, and processed by the publicly available
Fermi Science Tools version 9. 
The v9r15p2 software version and the P6\_V3\_DIFFUSE instrument response functions have been used. The light curve has been produced by an aperture photometry selecting photons from a region with radius 0.5 deg around 
the nominal position of PKS~1830-211 and energies between 300 MeV and 300 GeV.

%________________________________________________________________
\section{Data processing and method}

\subsection{Idea}
\label{idea}

If a distant source (in our case an AGN) is gravitationally lensed, 
the light reaches the observer through at least 2 different paths.
We assume here that there are only 2 light paths. 
In reality,  
the light curves of the 2 images are not totally
identical since (in addition to differences due to photon noise) 
the source can be microlensed in one of the 
two paths. In this paper, we neglect this effect.
Neglecting for the moment the background light, the observed flux can be decomposed into 2 components.
The first component is the intrinsic AGN light curve, given by $f(t),$ with Fourier transform $\tilde{f}(\nu ).$ 
The second component has a similar time evolution than the first one, but is shifted in time with a delay $a.$ 
In addition, 
the brightness of the second component  differs by a factor of $b$ 
from that of the first component, 
so that it can be written as $b f(t+a)$ and its transform to the Fourier space gives  $b\tilde{f}(\nu ) e^{-2\pi i\nu a}$.

The sum of two component gives
 $g(t) =  f(t) + bf(t + a)$ 
 which transforms into 
    $\tilde{g}(\nu ) = \tilde{f}(\nu ) (1 + b e^{-2\pi i\nu a})$
in Fourier space.

The power spectrum  $P_{\nu}$ of the source is obtained by computing the square
modulus of $\tilde{g}(\nu ).$

  \begin{equation}
      P_{\nu} = |\tilde{g}(\nu )|^{2} = |\tilde{f}(\nu )|^{2}(1 + b^{2} + 2b cos(2\pi \nu a)) 
\label{Powereq}
  \end{equation}

The measurable  $P_{\nu}$ is the product of the ``true'' power spectrum of the source times a periodic component 
with a period (in the frequency domain) equal to the inverse of the relative time delay $a$. 

The usual way of measuring the time delay $a$ is 
to calculate the autocorrelation 
function of f(t). This method was investigated by \citet{1996Geiger}.
The autocorrelation function can be written as the sum of three terms. 
One of this term describes the intrinsic autocorrelation of the AGN, 
decreasing with a time constant $\lambda$.
If $\lambda$ is larger than $a$, the autocorrelation method fails, 
because the time delay peak merges with the intrinsic component of the AGN.
Another potential problem with the autocorrelation method is the sensitivity 
to spurious periodicities such as the one coming from the motion and 
rotation of the Fermi satellite. 

The periodic modulation of $P_{\nu}$ suggests the use of another method, 
based on the computation of the power spectrum of $ P_{\nu} $, noted $D_{a}$.
This method is similar in spirit to the cepstrum analysis \citep{Bogert1963} used in seismology and speech processing.
If $|\tilde{f}(\nu )|^{2}$ was a constant function of $\nu$, $D_{a}$ would 
have a peak at the time delay $a$. 
In the general case, $D_{a}$ is obtained by the convolution of a Dirac function, coming from the cosinus modulation, 
by the Fourier transform of the function:

\begin{equation}
         \tilde{h}(\nu) = \tilde{f}(\nu )  Window(0,W) 
\end{equation}

where Window(0,W) is the window in frequency of $P_{\nu}$ and W is maximum available frequency. 
The Fourier transform h(a) of $ \tilde{h}(\nu)$ defines the width of the time delay peak
in the double power spectrum $D_{a}$.

For instance if $\tilde{f}(\nu) = e^{-\lambda\nu}$ and $\lambda W >> 1 ,$ 
then the time delay peak in $D_{a}$ has a Lorentzian shape  with a FWHM of ${\lambda}/{\pi}.$ 
For a typical value of $\lambda = 10$ days, one has a FWHM of 3 days.

In next section we describe the calculation of $P_{\nu}$ and $D_{a}$ and 
illustrate the procedure with Monte Carlo simulations. 

\subsection{Power Spectrum}
\label{FPSsubs}
The Fourier transform is a powerful technique to analyze astronomical data, 
but it requires a proper preparation of the observations. To avoid problems arising from the finite length of measurements,
 sampling and aliasing, we use the procedure for data reduction described by \citet{1971Brault}.

We start with dividing the whole light curve into several segments of equal length. The data have to be evenly spaced 
and the number of points per segment needs to be equal to a power of 2. 
The choice of the segment length is compromise between the spectral resolution and the size of error bars on points 
in the spectrum. The resolution of lines increases with the number of points in the segment, 
but the error on each point in the spectrum decreases as the number of segments.  

As suggested by \citet{1971Brault}, the segments are overlapped to obtain a larger number of segments 
with a sufficient number of  points.
 Then the data in each segment are transformed with the following procedure:
\begin{enumerate}
 \item Data gaps are removed by linear interpolation
 \item The mean is subtracted from the series to avoid having a large value in the first bin of the transform
 \item The data are oversampled
to remove aliasing. 
 \item Zeros are added to the end of the series to reduce power at high frequencies
\end{enumerate}
 
%----------------------------------------------------------- mc_fps-----------
   \begin{figure}
   \centering
 \includegraphics[width=8cm,angle=0]{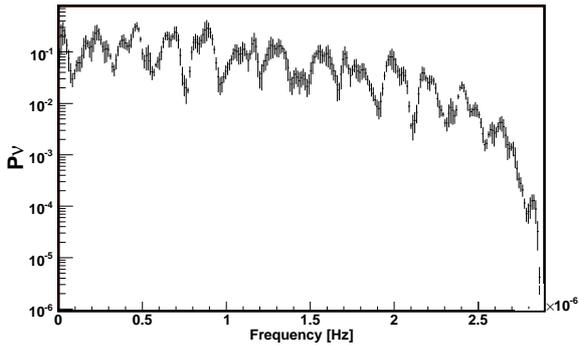} \\
 \caption{Simulated power spectrum of a lensed AGN with a time delay between images of 28 days. The power $P_{\nu}$ is in arbitrary units. } 
         \label{mc_fps}
   \end{figure}
%------------------------------------------------------------------------------

Finally, the power spectrum is averaged over all segments.
The power spectrum obtained with an artificial light curve is shown on figure \ref{mc_fps}.

The artificial light curve was produced by summing three simulated components.
The light curve of PKS 1830-211 shown on figure \ref{lc} 
 does not exhibit any easily recognizable features, but  has a rather random-like aspect.
The first component was thus simulated as a white noise, with a Poisson distribution. 
It would be more realistic to use red noise instead of white noise but the latter is sufficient for most of our purposes, 
such as computing $D_{a}.$ The second component is obtained from the first by shifting the light curve 
with a 28 days time lag. The effect of differential magnification of the images has also been included.
The background photon noise was taken into account by adding a third component 
with a Poisson distribution.

The mean number of counts per 2 day bin for PKS 1830-211 is 5.42 counts. 
This value was used  for the simulation of the artificial light curve.
The first and second component contribute 80\% of the simulated count rate 
and the rest is contributed by the noise component.

To get a simulated flux similar to the observed flux, the simulated count 
rate is divided by 
the average exposure from observations ( $2.8557\ 10^{7}\ s\ cm^{2}$).

\subsection{Time Delay Determination}
\label{SubTDD}
 
The methods of time delay determination use the power spectrum $P_{\nu}$ obtained as described in the previous section. 
The simulated $P_{\nu}$ presented on figure \ref{mc_fps} shows a very clear periodic pattern. 
From equation \ref{Powereq} we know that the period of the observed wobbles corresponds 
to inverse of the time delay between the images. 

Our preferred approach was to calculate the double power spectrum $D_{a}$.  
As in section \ref{FPSsubs}, the power spectrum $P_{\nu}$ has to be prepared 
before undergoing a Fourier transform to the ``time delay'' domain.
 
\begin{figure}
   \centering
 \includegraphics[width=8cm,angle=0]{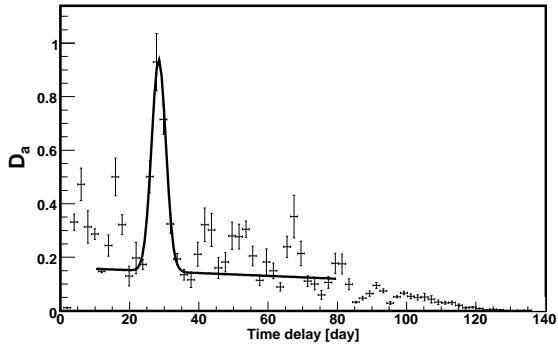} \\
 \caption{Double power spectrum $D_{a}$ for the simulated lensed AGN of figure \ref{mc_fps}. $D_{a}$ is plotted in arbitrary units.
          The solid line is a fit to a linear plus Gaussian profile. } 
         \label{mc_sps}
   \end{figure}

The low frequency part ($\nu < 1/55 \mbox{day}^{-1}$) 
of $P_{\nu}$ is cut off. This cut arises because 
of the large power observed low frequencies in the power spectrum 
of PKS~1830-211.  
The high frequency part  of the spectrum $P_{\nu}$ is also removed because 
the power at high frequency is small (it goes to 0 at the Nyquist frequency).      
The calculation of $D_{a}$ proceeds like in section \ref{FPSsubs}, 
except that the $P_{\nu}$ data are bend to zero by multiplication with a cosine bell. 
This eliminates spurious high frequencies, when zeros are added to the $P_{\nu}$ series.
The $D_{a}$ distribution is estimated from 5 segments of the light curve. 
In every bin of the  $D_{a}$ distribution, the estimated double power spectrum 
is given by the average over the 5 segments.
The errors bars on $D_{a}$ are estimated from the dispersion of bin values divided by 2 (since there are 5 segments). 
Due to the random nature of the sampling process, some of the error bars obtained 
are much smaller than the typical dispersion in the  $D_{a}$ points. 
To take this into account, a small systematic error bar was added quadratically to all points.
The result (with statistical error bars only) is presented on 
figure \ref{mc_sps}.

As described in section \ref{FPSsubs}, we simulated light curves with a time delay of 28 days.
A peak is apparent near a time delay of 28 days on the $D_{a}$ distribution
shown on figure \ref{mc_sps}. 
The points just outside the peak are compatible with a flat distribution.
Including also the points in the peak gives a distribution which is incompatible with a flat distribution at the 10 sigma level. 
 
The parameters of the peak were determined by fitting the sum of a linear function for the background plus a Gaussian function for the signal.  
In the case shown on figure \ref{mc_sps}, the time delay estimated from $D_{a}$ is $27.94\pm0.61$ days.

\begin{figure}
   \centering
 \includegraphics[width=8cm,angle=0]{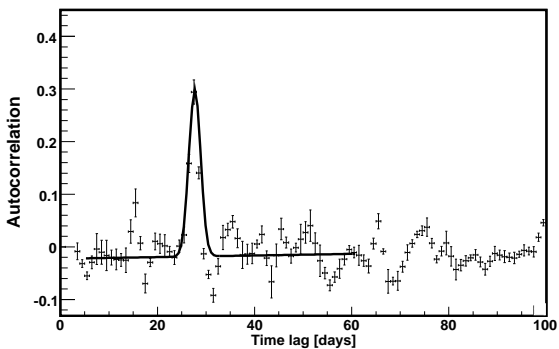} \\
 \caption{Autocorrelation function of the simulated lensed AGN of figure \ref{mc_fps}.
          The solid line is a fit to a linear plus Gaussian profile.} 
         \label{mc_ac}
   \end{figure}

As mentioned in section \ref{idea}, the usual approach for the time delay estimation 
is to compute the autocorrelation of the light curve. 
The auto-covariance is obtained by taking the real part of the inverse Fourier transform of $P_{\nu}$.
The auto-covariance is normalized (divided by the value at zero time lag) to get the autocorrelation.
The autocorrelation function of an artificial light curve simulated as in section \ref{FPSsubs} is presented on figure \ref{mc_ac}. 

A peak with a significance of roughly 16 $\sigma$ is present at  $27.85\pm0.14$ days. 
However the significance of this peak is overestimated since 
light curves are simulated with white noise instead of red noise.
The autocorrelation function of a light curve driven by red noise is given by  $e^{-a/ \lambda}.$
In the case of our simulated light curves, $\lambda = 0$, so that the peak is a little affected by the background
of the AGN. 

For the simulated light curves, both approaches of time delay determination give reasonable and compatible results.

%----------------------------------------------------------- mc_sps-----------
%-----------------------------------------------------------------------------

%----------------------------------------------------------- mc_ac-----------
%-----------------------------------------------------------------------------

%______________________________________________________________________________
\section{Results and Discussion}

\begin{figure}
   \centering
 \includegraphics[width=8cm,angle=0]{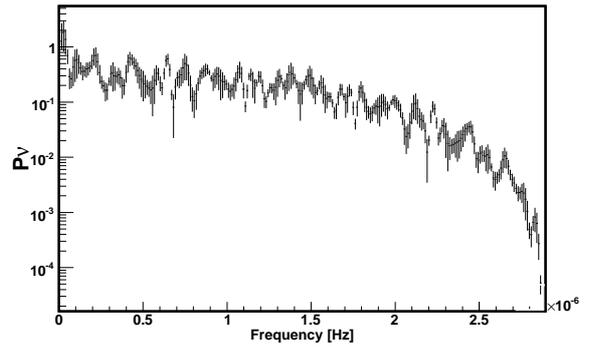} \\
 \caption{ Measured power spectrum of PKS 1830-211, plotted in arbitrary units.  } 
         \label{pks1830_fps}
   \end{figure}

The results for real data were obtained with the same procedure as was presented for the simulated light curves.
Figure \ref{pks1830_fps} shows the power spectrum  $P_{\nu}$ calculated from the light curve of PKS 1830-211.
A periodic pattern is obvious on the distribution of $P_{\nu}$  calculated 
from the PKS 1830-211 light curve. 
It is similar to the pattern expected from the simulations shown on figure \ref{mc_fps}. 
  
\begin{figure}
   \centering
 \includegraphics[width=8cm,angle=0]{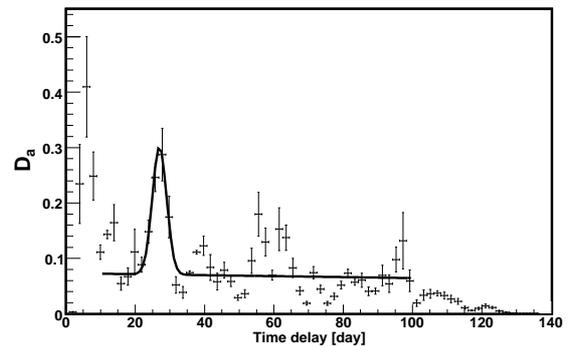} \\
 \caption{Double power spectrum of PKS 1830-211 plotted in arbitrary units. 
          The solid line is a fit to a linear plus Gaussian profile. } 
         \label{pks1830_sps}
   \end{figure}

\begin{figure}
   \centering
 \includegraphics[width=8cm,angle=0]{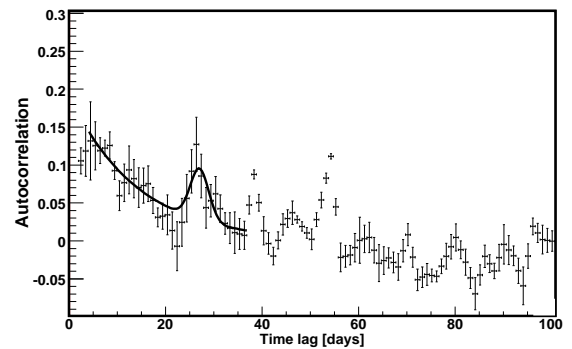} \\
 \caption{Measured autocorrelation function of PKS 1830-211. The solid line is a fit to an exponential plus Gaussian profile.} 
         \label{pks1830_ac}
   \end{figure}

The autocorrelation function and the $D_{a}$ distribution calculated for real data 
are shown on figures \ref{pks1830_ac} and \ref{pks1830_sps}. 
A peak around 27 days is seen in both distributions.
Several other peaks are present on the autocorrelation function as was already noted by \citet{1996Geiger} (see their Fig 1).
The peak around 5 days in the $D_{a}$ distribution is likely to be an
artefact connected to the time variation of the exposure of the LAT instrument
on PKS 1830-211.
Using the method described in section \ref{SubTDD}, the significance of the peak around 27 days is found to be 1.1 $\sigma$ 
in the autocorrelation function and 3 $\sigma$ in the double power spectrum $D_{a}$. 
Fitting the position of the peak gives the time delay of $a= 27.5\pm1.3$ days for the $D_{a}$ distribution.
The fit of the autocorrelation function to a gaussian peak over an exponential background gives $a =  27.07\pm0.45$ days. 

The double power spectrum distribution obtained for PKS 1830-211 provides the first evidence for gravitational lensing
phenomena in high energy gamma rays.
The evidence is still at the 3 $\sigma$ level but will likely improve by a factor of 2 over the lifetime of the Fermi satellite.   
Thanks to the uniform light curve sampling provided by Fermi LAT instrument,
it is not necessary to identify features on the light curve to apply Fourier transform methods.
The example of PKS 1830-211 shows that the method works in spite of the low photon statistic.
Possible extensions of the present work are finding multiple delays in complicated lens systems
or looking for unknown lensing systems in the Fermi catalog of AGNs.

%-----------------------------------------------------------Figure: pks 1830 fps-----------
%------------------------------------------------------------------------------

%-----------------------------------------------------------Figure: pks 1830 sps-----------
%------------------------------------------------------------------------------

%-----------------------------------------------------------Figure: pks 1830 Auto Correlation-----------
%------------------------------------------------------------------------------

\bibliographystyle{aa}
\bibliography{pks1830}

\begin{thebibliography}{17}
\expandafter\ifx\csname natexlab\endcsname\relax\def\natexlab#1{#1}\fi

\bibitem[{{Abdo} {et~al.}(2010){Abdo}, {Ackermann}, {Ajello}, {Allafort},
  {Antolini}, {Atwood}, {Axelsson}, {Baldini}, {Ballet}, {Barbiellini},
  {Bastieri}, {Baughman}, {Bechtol}, {Bellazzini}, {Belli}, {Berenji},
  {Bisello}, {Blandford}, {Bloom}, {Bonamente}, {Bonnell}, {Borgland},
  {Bouvier}, {Bregeon}, {Brez}, {Brigida}, {Bruel}, {Burnett}, {Busetto},
  {Buson}, {Caliandro}, {Cameron}, {Campana}, {Canadas}, {Caraveo}, {Carrigan},
  {Casandjian}, {Cavazzuti}, {Ceccanti}, {Cecchi}, {Celik}, {Charles},
  {Chekhtman}, {Cheung}, {Chiang}, {Cillis}, {Ciprini}, {Claus},
  {Cohen-Tanugi}, {Conrad}, {Corbet}, {Davis}, {Deklotz}, {den Hartog},
  {Dermer}, {de Angelis}, {de}, {de Palma}, {Digel}, {Dormody}, {Do Couto},
  {Drell}, {Dubois}, {Dumora}, {Fabiani}, {Farnier}, {Favuzzi}, {Fegan},
  {Ferrara}, {Focke}, {Fortin}, {Frailis}, {Fukazawa}, {Funk}, {Fusco},
  {Gargano}, {Gasparrini}, \& {Gehrels}}]{2010Abdo}
{Abdo}, A.~A., {Ackermann}, M., {Ajello}, M., {et~al.} 2010, VizieR Online Data
  Catalog, 218, 80405

\bibitem[{{Atwood} {et~al.}(2009){Atwood}, {Abdo}, {Ackermann}, {Althouse},
  {Anderson}, {Axelsson}, {Baldini}, {Ballet}, {Band}, {Barbiellini}, \&
  et~al.}]{2009Atwood}
{Atwood}, W.~B., {Abdo}, A.~A., {Ackermann}, M., {et~al.} 2009, \apj, 697, 1071

\bibitem[{{Bogert} {et~al.}(1963){Bogert}, {Healy}, \& {Tukey}}]{Bogert1963}
{Bogert}, B.~P., {Healy}, M.~J.~R., \& {Tukey}, J.~W. 1963, in Proceedings on
  the Symposium on Time Series Analysis, ed. {M.Rosenblat}, Wiley, NY, 209--243

\bibitem[{{Bolton} {et~al.}(2004){Bolton}, {Burles}, {Schlegel}, {Eisenstein},
  \& {Brinkmann}}]{2004Bolton}
{Bolton}, A.~S., {Burles}, S., {Schlegel}, D.~J., {Eisenstein}, D.~J., \&
  {Brinkmann}, J. 2004, \aj, 127, 1860

\bibitem[{{Brault} \& {White}(1971)}]{1971Brault}
{Brault}, J.~W. \& {White}, O.~R. 1971, \aap, 13, 169

\bibitem[{{Browne} {et~al.}(2003){Browne}, {Wilkinson}, {Jackson}, {Myers},
  {Fassnacht}, {Koopmans}, {Marlow}, {Norbury}, {Rusin}, {Sykes}, {Biggs},
  {Blandford}, {de Bruyn}, {Chae}, {Helbig}, {King}, {McKean}, {Pearson},
  {Phillips}, {Readhead}, {Xanthopoulos}, \& {York}}]{2003Browne}
{Browne}, I.~W., {Wilkinson}, P.~N., {Jackson}, N.~J., {et~al.} 2003, \mnras,
  341, 13

\bibitem[{{Geiger} \& {Schneider}(1996)}]{1996Geiger}
{Geiger}, B. \& {Schneider}, P. 1996, \mnras, 282, 530

\bibitem[{{Jauncey} {et~al.}(1991){Jauncey}, {Reynolds}, {Tzioumis}, {Muxlow},
  {Perley}, {Murphy}, {Preston}, {King}, {Patnaik}, {Jones}, {Meier}, {Bird},
  {Blair}, {Bunton}, {Clay}, {Costa}, {Duncan}, {Ferris}, {Gough}, {Hamilton},
  {Hoard}, {Kemball}, {Kesteven}, {Lobdell}, {Luiten}, {Mcculloch}, {Murray},
  {Nicholson}, {Rao}, {Savage}, {Sinclair}, {Skjerve}, {Taaffe}, {Wark}, \&
  {White}}]{1991Jauncey}
{Jauncey}, D.~L., {Reynolds}, J.~E., {Tzioumis}, A.~K., {et~al.} 1991, \nat,
  352, 132

\bibitem[{{Lidman} {et~al.}(1999){Lidman}, {Courbin}, {Meylan}, {Broadhurst},
  {Frye}, \& {Welch}}]{1999Lidman}
{Lidman}, C., {Courbin}, F., {Meylan}, G., {et~al.} 1999, \apjl, 514, L57

\bibitem[{{Lovell} {et~al.}(1998){Lovell}, {Jauncey}, {Reynolds}, {Wieringa},
  {King}, {Tzioumis}, {McCulloch}, \& {Edwards}}]{1998Lovell}
{Lovell}, J.~E.~J., {Jauncey}, D.~L., {Reynolds}, J.~E., {et~al.} 1998, \apjl,
  508, L51

\bibitem[{{Lovell} {et~al.}(1996){Lovell}, {Reynolds}, {Jauncey}, {Backus},
  {McCulloch}, {Sinclair}, {Wilson}, {Tzioumis}, {King}, {Gough}, {Ellingsen},
  {Phillips}, {Preston}, \& {Jones}}]{1996Lovell}
{Lovell}, J.~E.~J., {Reynolds}, J.~E., {Jauncey}, D.~L., {et~al.} 1996, \apjl,
  472, L5+

\bibitem[{{Mattox} {et~al.}(1997){Mattox}, {Schachter}, {Molnar}, {Hartman}, \&
  {Patnaik}}]{1997Mattox}
{Mattox}, J.~R., {Schachter}, J., {Molnar}, L., {Hartman}, R.~C., \& {Patnaik},
  A.~R. 1997, \apj, 481, 95

\bibitem[{{Myers} {et~al.}(2003){Myers}, {Jackson}, {Browne}, {de Bruyn},
  {Pearson}, {Readhead}, {Wilkinson}, {Biggs}, {Blandford}, {Fassnacht},
  {Koopmans}, {Marlow}, {McKean}, {Norbury}, {Phillips}, {Rusin}, {Shepherd},
  \& {Sykes}}]{2003Myers}
{Myers}, S.~T., {Jackson}, N.~J., {Browne}, I.~W.~A., {et~al.} 2003, \mnras,
  341, 1

\bibitem[{{Pramesh Rao} \& {Subrahmanyan}(1988)}]{1988Rao}
{Pramesh Rao}, A. \& {Subrahmanyan}, R. 1988, \mnras, 231, 229

\bibitem[{{Torres} {et~al.}(2003){Torres}, {Romero}, {Eiroa}, {Wambsganss}, \&
  {Pessah}}]{2003Torres}
{Torres}, D.~F., {Romero}, G.~E., {Eiroa}, E.~F., {Wambsganss}, J., \&
  {Pessah}, M.~E. 2003, \mnras, 339, 335

\bibitem[{{Wiklind} \& {Combes}(1996)}]{1996Wiklind}
{Wiklind}, T. \& {Combes}, F. 1996, in Astrophysics and Space Science Library,
  Vol. 206, Cold Gas at High Redshift, ed. {M.~N.~Bremer \& N.~Malcolm}, 227--+

\bibitem[{{Winn} {et~al.}(2002){Winn}, {Kochanek}, {McLeod}, {Falco}, {Impey},
  \& {Rix}}]{2002Winn}
{Winn}, J.~N., {Kochanek}, C.~S., {McLeod}, B.~A., {et~al.} 2002, \apj, 575,
  103

\end{thebibliography}

\end{document}